\documentclass{appolb}
\usepackage{epsfig}
\usepackage{wrapfig}   

\begin{document}
\title{News from strong interactions program of the NA61/SHINE experiment  %
\thanks{Presented at the Critical Point and Onset of Deconfinement 2016, Wroclaw, Poland, May 30th - June 4th, 2016 }%
}
\author{Katarzyna Grebieszkow for the NA61/SHINE Collaboration 
\address{Faculty of Physics, Warsaw University of Technology,
Koszykowa 75, 00-662~Warsaw, Poland}
}
\maketitle
\begin{abstract}
The NA61/SHINE experiment aims to discover the critical point of strongly interacting matter and study the properties of the onset of deconfinement. This is performed by a two dimensional phase diagram ($T-\mu_B$) scan of measurements of particle spectra and fluctuations in proton-proton, proton-nucleus and nucleus-nucleus interactions as a function of collision energy and system size. 
In this contribution new NA61/SHINE results on negative pion production, as well as transverse momentum and multiplicity fluctuations in Ar+Sc collisions are presented. Moreover, the latest results on higher order moments of net-charge multiplicity distribution in p+p collisions are also discussed. The Ar+Sc results are compared to NA61 p+p and Be+Be data, as well as to NA49 $A+A$ results. 
\end{abstract}


\section{Introduction}
NA61/SHINE at the CERN Super Proton Synchrotron (SPS) is a fixed-target experiment pursuing a rich physics program including measurements for strong interactions, neutrino, and cosmic ray physics. The main goal of the strong interactions program is to study the characteristics of the onset of deconfinement and search for the signatures of the critical point (CP). Presently, there are several more projects at energies corresponding to the SPS energy range: the RHIC Beam Energy Scan (BNL, Brookhaven), NICA (JINR, Dubna), and SIS-100/300 (FAIR GSI, Darmstadt). NA61 is the first experiment to perform a two-dimensional scan, in beam momentum and mass number of colliding nuclei. 

Contrary to collider experiments, the acceptance of NA61 starts from $p_T=0$ MeV/c. This is crucial not only for the analysis of spectra and yields, but also for the study of fluctuations, because it is expected that fluctuations due to the CP mainly show up in low $p_T$ particles \cite{Stephanov:1999zu}. As will be shown below, the rapidity coverage of NA61 is also very large. It essentially covers the full forward hemisphere and extends below mid-rapidity allowing to determine particle multiplicities in full phase-space (at RHIC only the BRAHMS experiment has been able to measure $4\pi$ multiplicities).  

In many experiments centrality selection is based on produced particle multiplicity. But multiplicity-based event selection may bias multiplicity fluctuations! To avoid such a bias centrality is measured in NA61 using the Projectile Spectator Detector (PSD), a segmented calorimeter recording the forward going energy $E_F$ with resolution of 1 nucleon in the studied energy range. The PSD is located on the beam axis and detects mainly the non-interacting nucleons of the beam nucleus. Intervals in $E_F$ allow to select different centrality classes. 
The PSD will also be used for determination of the reaction plane in the future analysis of collective flow. 

The analysis methods in NA61 are also unique and innovative. The {\it identity method} \cite{Gazdzicki:2011xz, Gorenstein:2011hr, Rustamov:2012bx}, allowing to unfold the effect of imperfect particle identification, was used to obtain identified particle spectra \cite{Pulawski:2015tka} and identified particle fluctuations \cite{Mackowiak-Pawlowska:2013caa}. Moreover, NA61 results, not only on spectra but also on fluctuations, are corrected for non-target interactions, detector inefficiencies, and trigger bias (see Refs.\cite{Abgrall:2013qoa, Aduszkiewicz:2015jna} for details). 

Finally, for fluctuation analysis numerous measures are proposed in the literature. However, it is very important to use those which allow to eliminate trivial sources of fluctuations and keep only the fluctuations of interest. NA61 therefore uses new {\it strongly intensive} measures of fluctuations which are not only properly normalized (see below) but also allow to get rid of trivial dependences on volume and volume fluctuations.

\section{Studying the properties of the onset of deconfinement}

One of the main reasons to analyze spectra and yields in NA61 is to study the properties of the onset of deconfinement by looking whether the {\it kink}, {\it horn}, and {\it step} \cite{Gazdzicki:1998vd} structures are present also in collisions of small and intermediate mass nuclei. For Pb+Pb interactions NA49 showed a sharp peak ({\it horn}) in the $K^{+}/\pi^{+}$ ratio interpreted as due to onset of deconfinement~\cite{Gazdzicki:2014sva}. Moreover, a plateau ({\it step}) in the inverse slope parameter ($T$) of $m_T$ spectra was also observed as expected for constant temperature and pressure in a mixed phase. The recent NA61 results \cite{Pulawski:2015tka} (see also this conference slides for $K^{+}/\pi^{+}$ ratio in $4\pi$ acceptance) show that even in p+p collisions the energy dependences of $K^{+}/\pi^{+}$ and $T$ exhibit rapid changes (however without a maximum for $K^{+}/\pi^{+}$) in the SPS energy range. In this contribution {\it kink} plots are shown with new preliminary NA61 results from Ar+Sc and Be+Be interactions.



\begin{figure}[ht]
\centering
\includegraphics[width=0.23\textwidth]{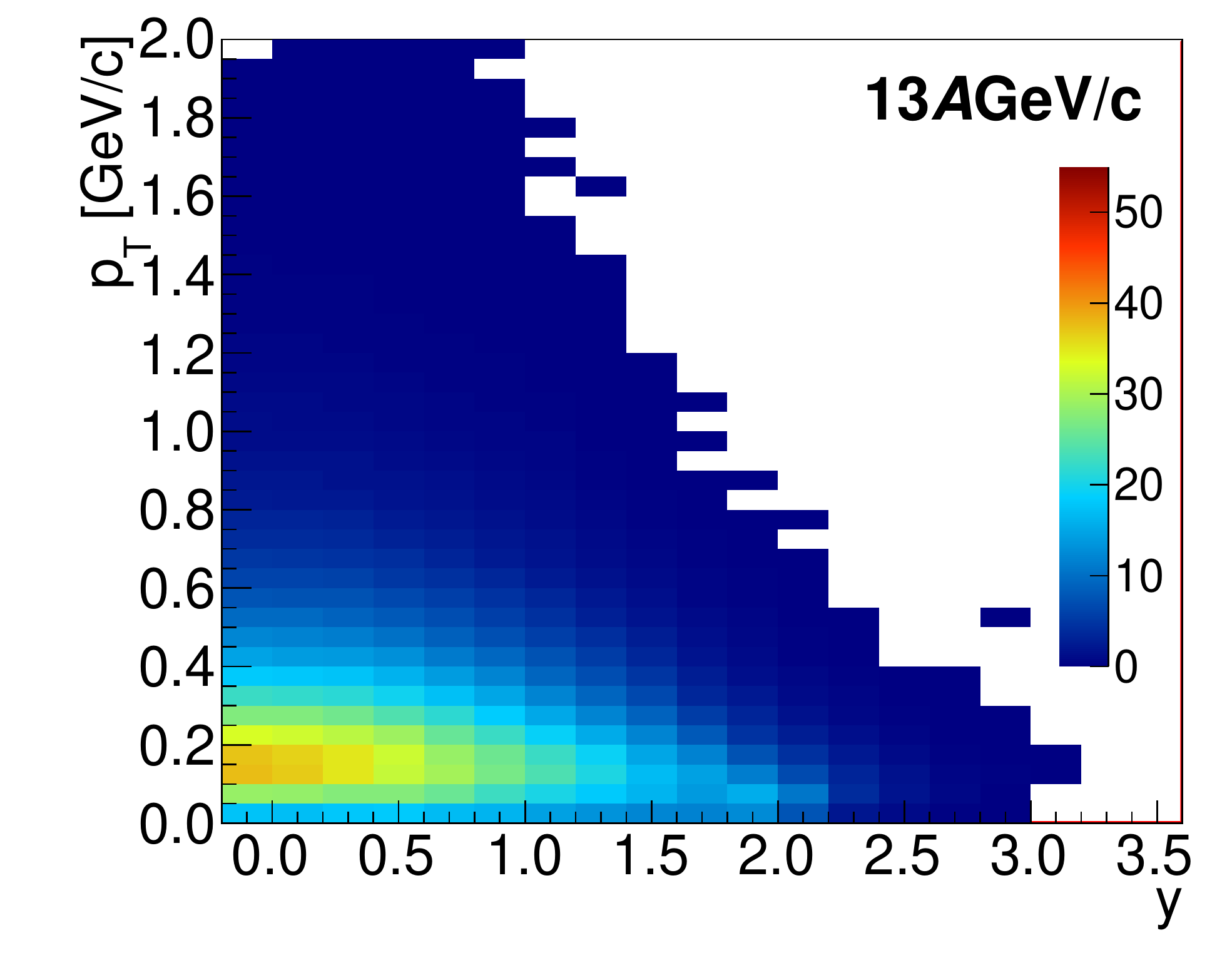}
\includegraphics[width=0.23\textwidth]{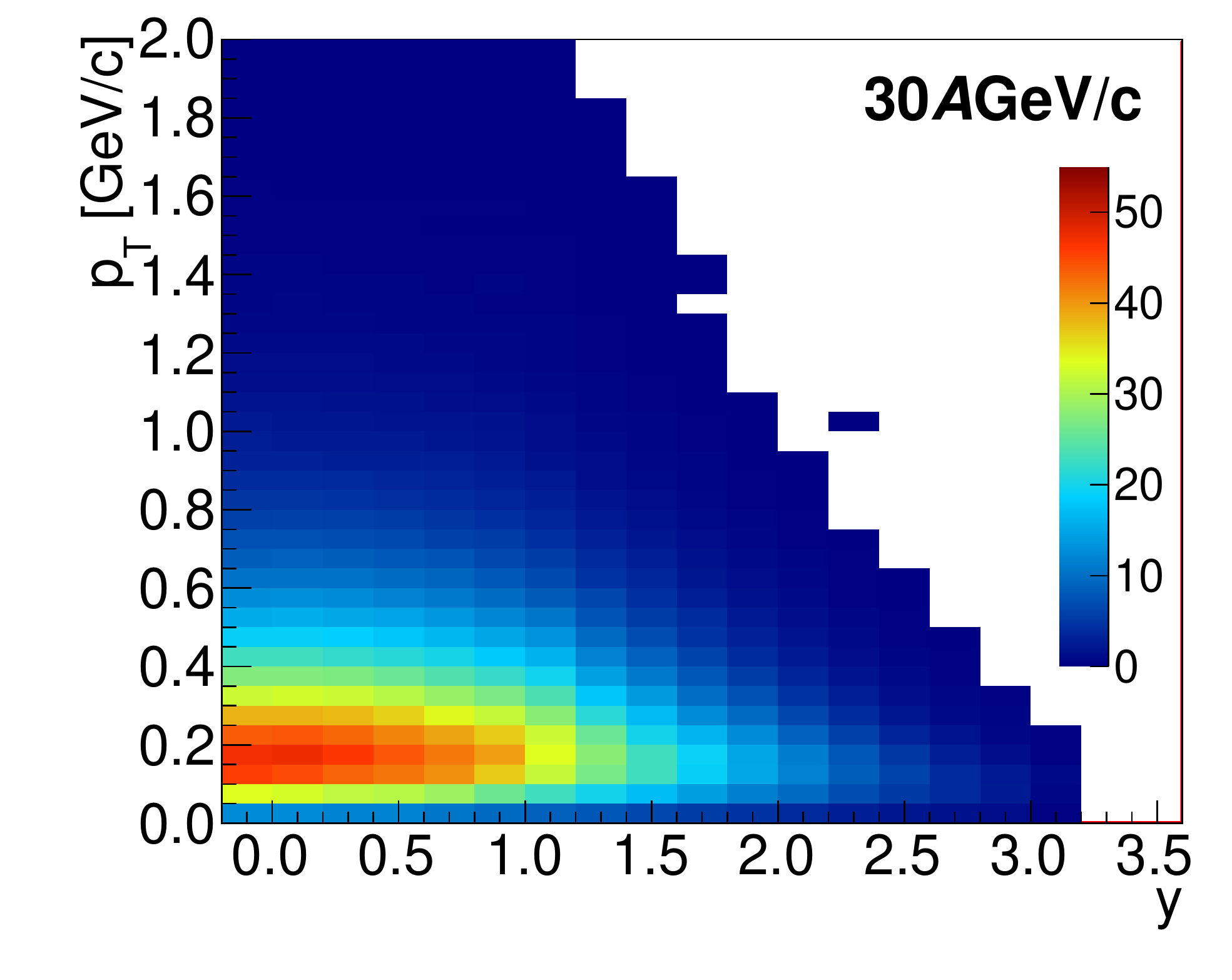}
\includegraphics[width=0.23\textwidth]{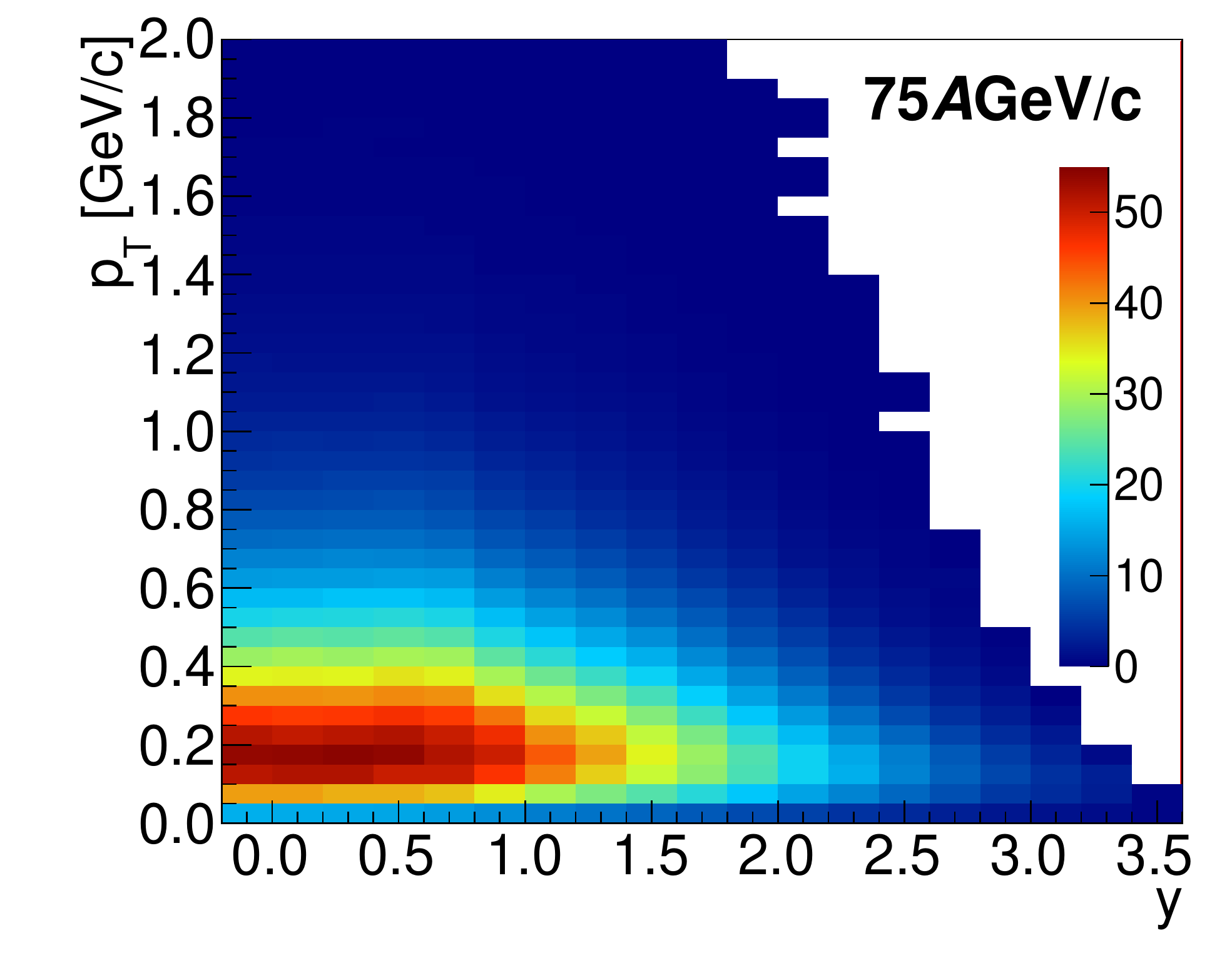}
\includegraphics[width=0.23\textwidth]{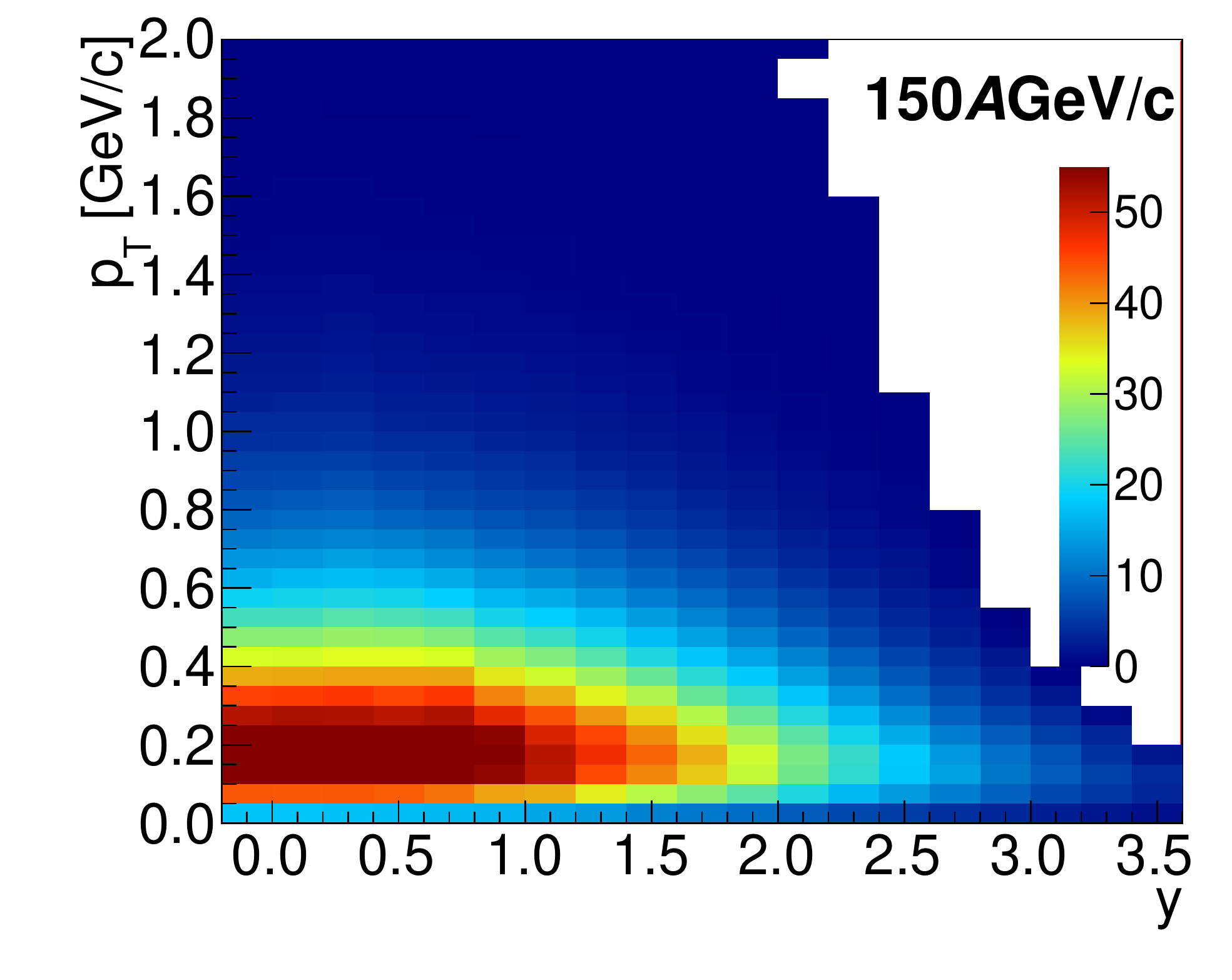}
\vspace{-0.3cm}
\caption[]{$d^{2}n/(dp_T dy)$ spectra of $\pi^{-}$ mesons in 0-5\% central Ar+Sc collisions.}
\label{2dspectra_arsc}
\end{figure}

The spectra of $\pi^{-}$ in Ar+Sc collisions were obtained using the so-called $h^{-}$ analysis method assuming that the majority of negatively charged particles are $\pi^{-}$ mesons. The contribution ($\approx10\% $) of other particles ($K^{-}$, $\overline{p}$) was subtracted using EPOS 1.99. 
Examples (see also Ref.~\cite{maciej_cpod16}) of double differential spectra of negatively charged pions in rapidity and transverse momentum for the most central 
Ar+Sc interactions are presented in Fig.~\ref{2dspectra_arsc}.

\begin{figure}[ht]
\centering
\includegraphics[width=0.23\textwidth]{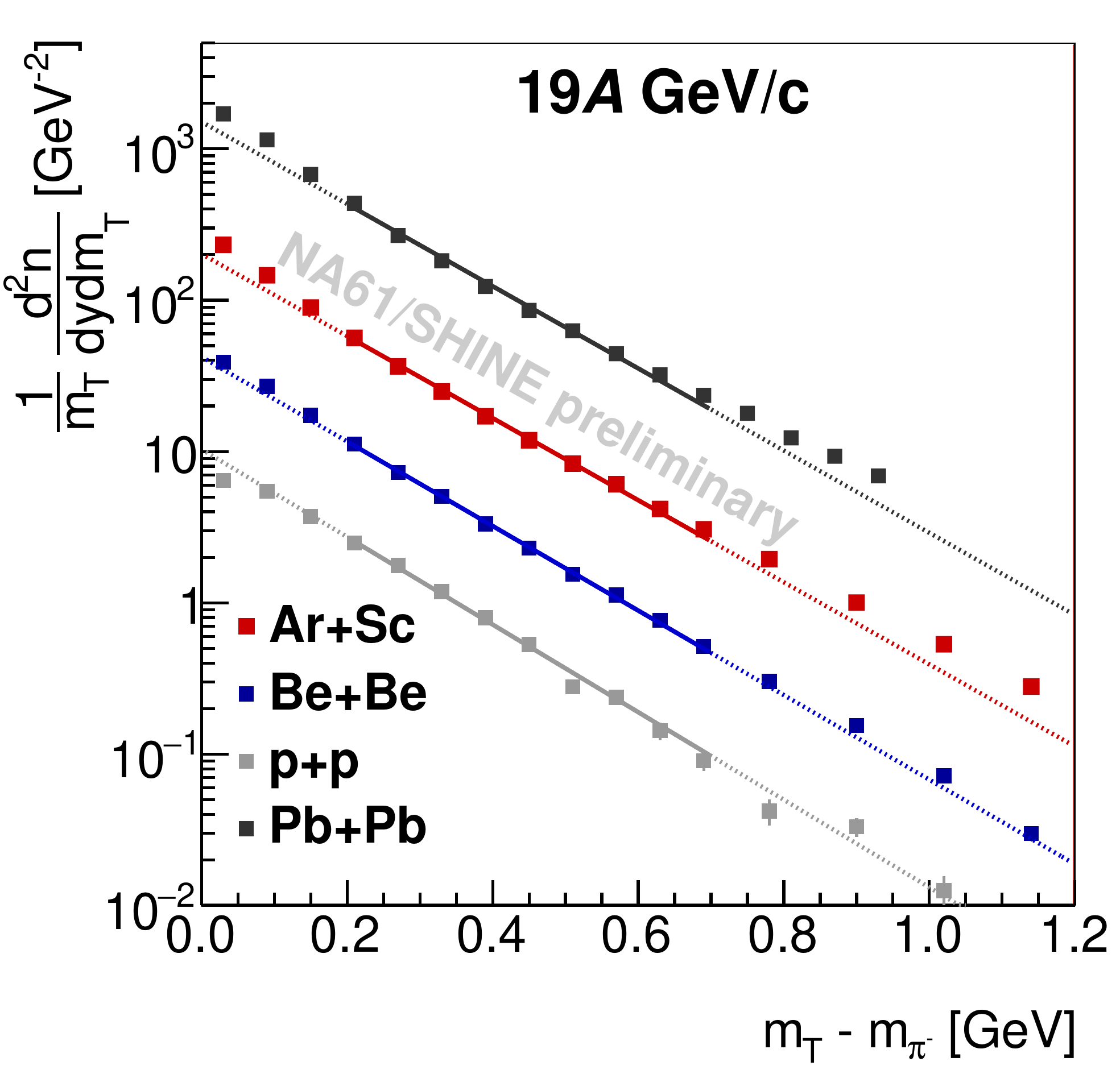}
\includegraphics[width=0.23\textwidth]{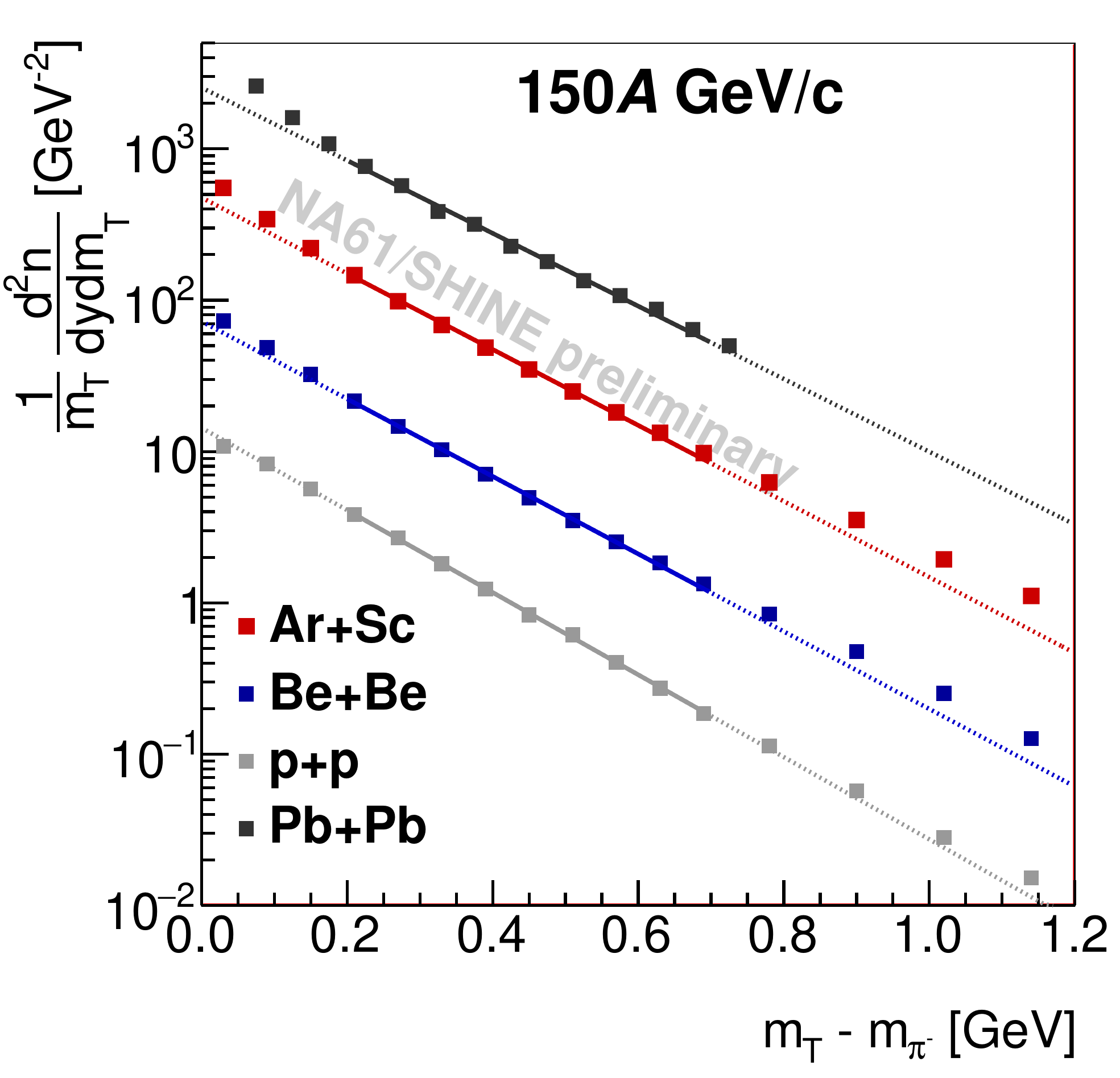}
\includegraphics[width=0.25\textwidth]{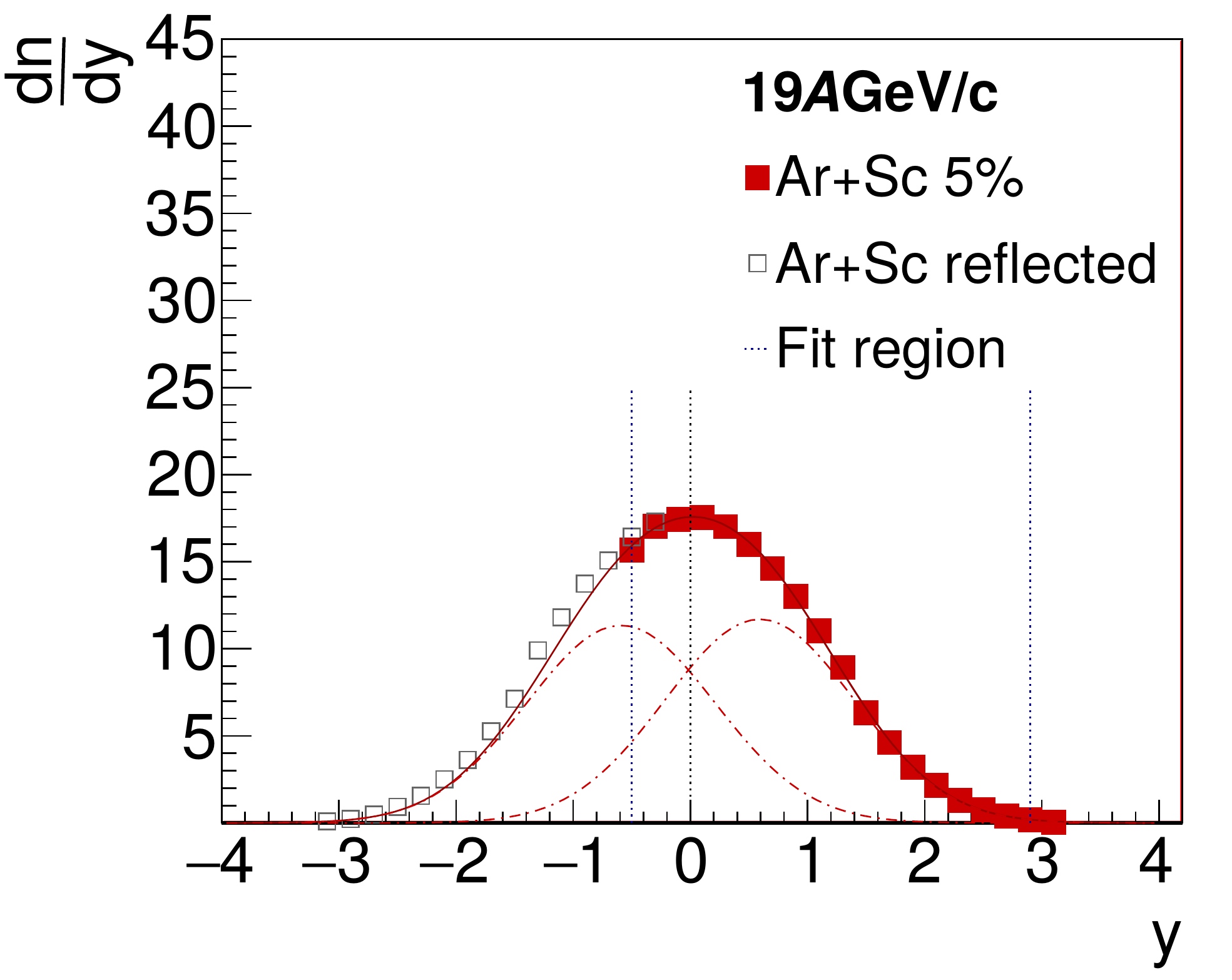}
\includegraphics[width=0.25\textwidth]{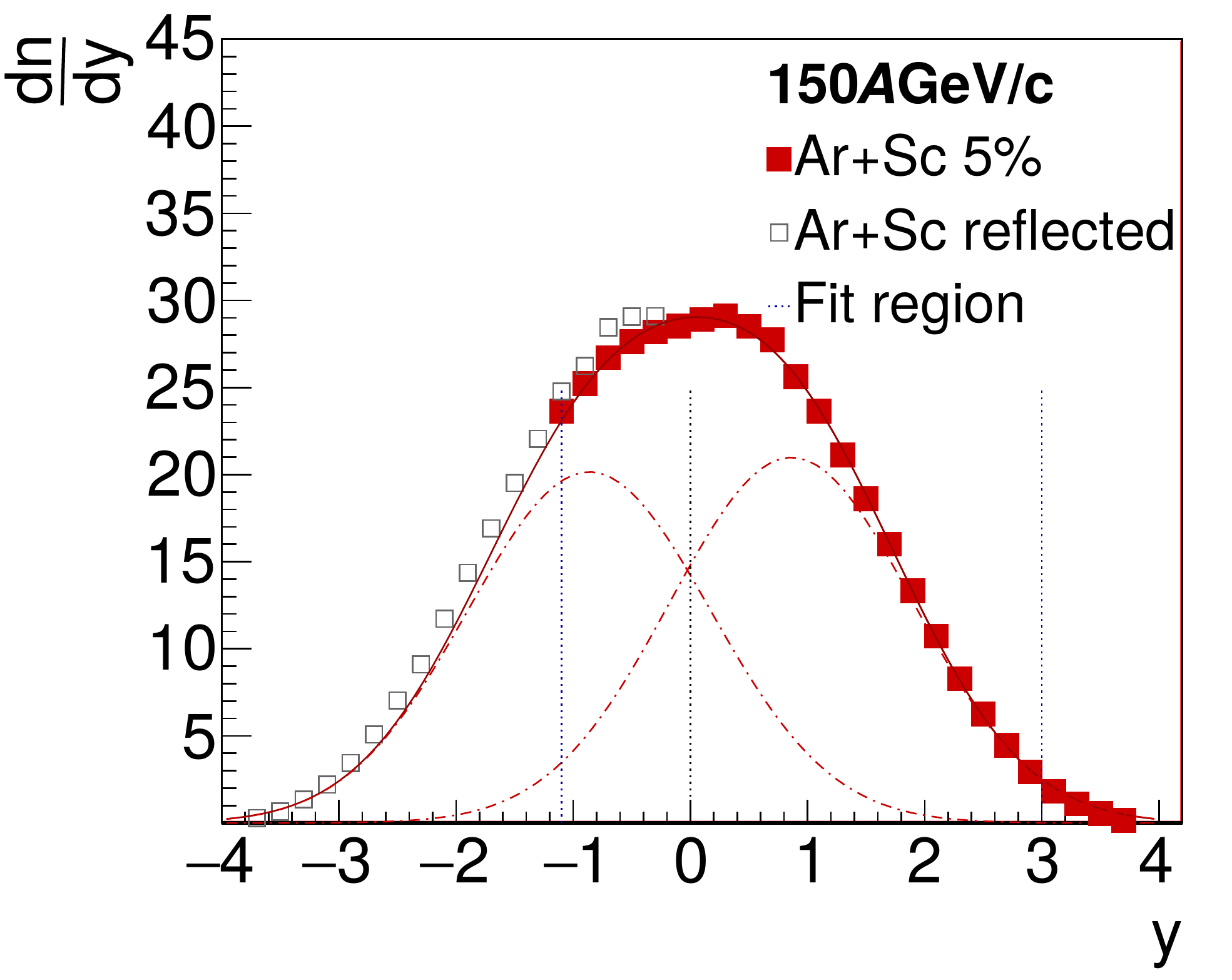}
\vspace{-0.2cm}
\caption[]{Transverse mass spectra (two left) and rapidity spectra (two right) of $\pi^{-}$ mesons in 0-5\% Ar+Sc collisions. The $m_T$ spectra are compared to NA49 Pb+Pb \cite{Afanasiev:2002mx, Alt:2007aa} and NA61 p+p \cite{Abgrall:2013qoa} and Be+Be results. Note: p+p and Pb+Pb collisions were taken at 158$A$ GeV/c beam momenta. }
\label{mt_rap_spectra_arsc}
\end{figure}

Some examples of preliminary $m_T$ spectra of $\pi^{-}$ mesons in Ar+Sc collisions are shown in Fig.~\ref{mt_rap_spectra_arsc} and compared to published NA61 p+p spectra \cite{Abgrall:2013qoa}, preliminary NA61 Be+Be data, and NA49 Pb+Pb results \cite{Afanasiev:2002mx, Alt:2007aa}. The p+p $m_T$ spectra are exponential, but of a different shape than seen for Pb+Pb, Ar+Sc, and Be+Be interactions, which may originate from isospin effects, collective flow, changing role of resonance production.

\begin{figure}[ht]
\centering
\vspace{-0.35cm}
\includegraphics[width=0.36\textwidth]{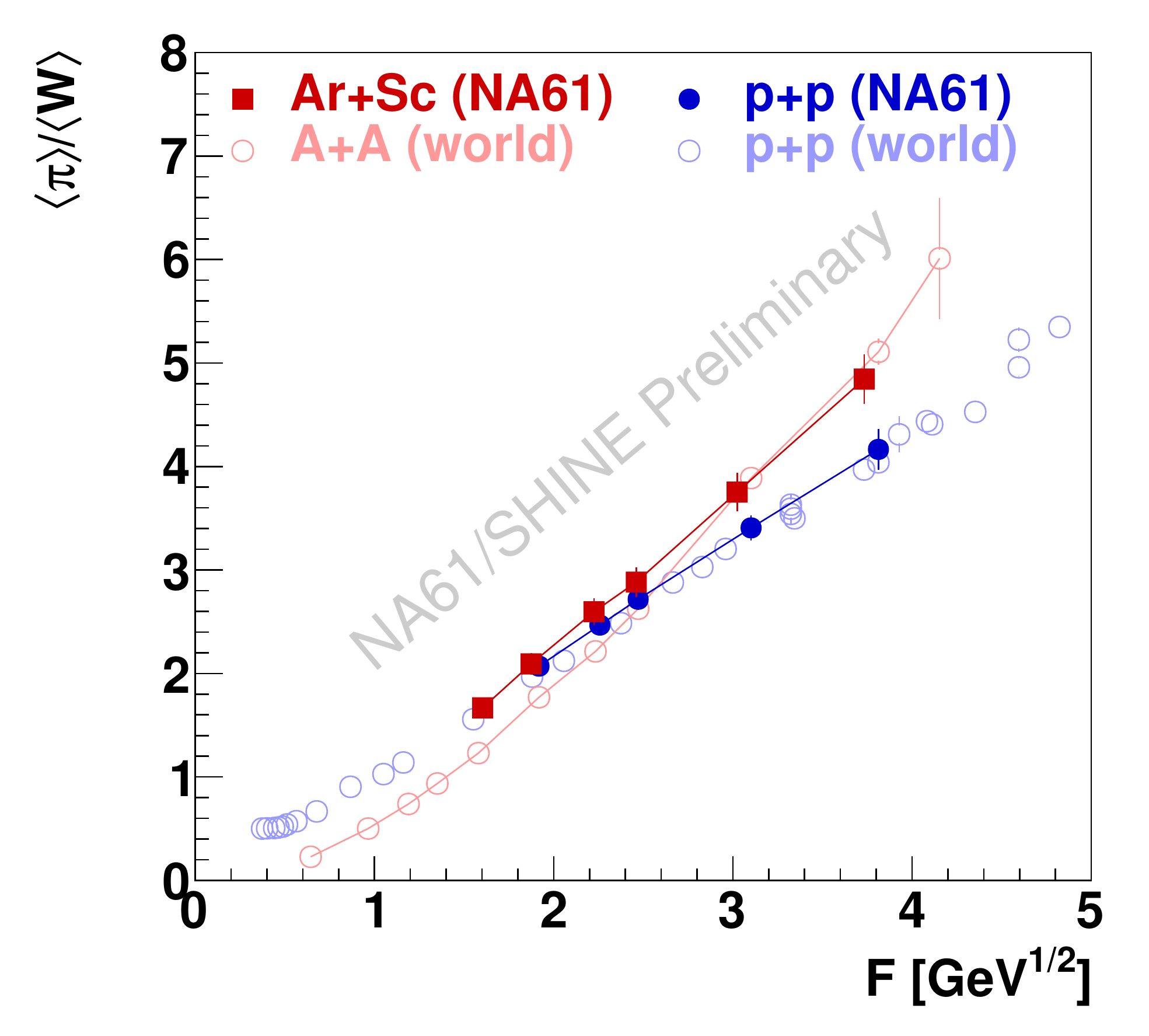}
\includegraphics[width=0.36\textwidth]{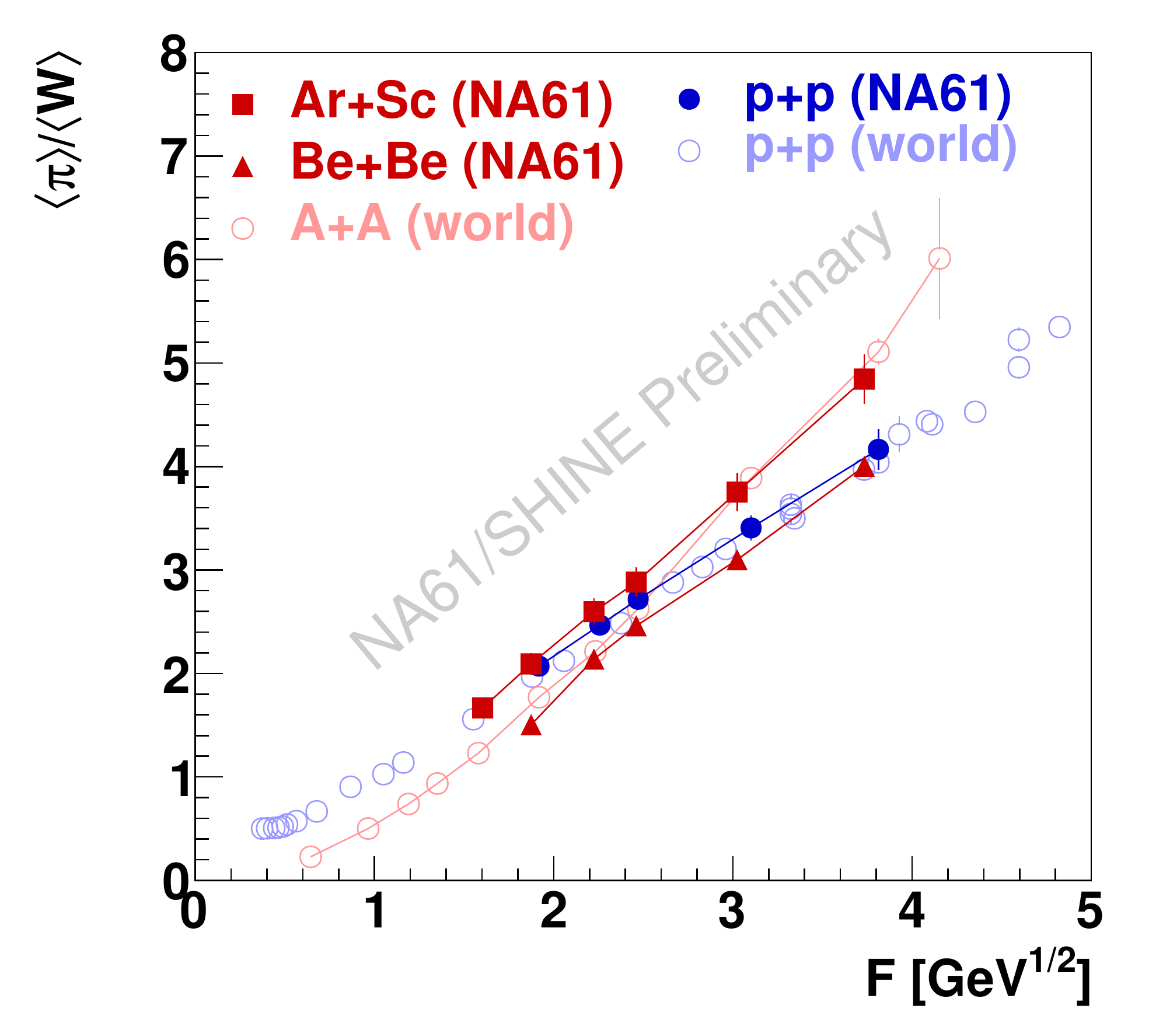}
\vspace{-0.3cm}
\caption[]{{\it Kink} plots: mean pion multiplicity divided by mean number of wounded nucleons versus Fermi energy $F$ measure \cite{Gazdzicki:1998vd} ($F \simeq s^{1/4}_{NN}$). See Ref.~\cite{michal_cpod16} for details of calculating ($4\pi$) mean pion multiplicity, $W$, and for references to word data.}
\label{kink_plot}
\end{figure}

The corresponding $p_T$-extrapolated and $p_T$-integrated rapidity spectra of $\pi^{-}$ mesons are shown in Fig.~\ref{mt_rap_spectra_arsc} (two right panels). One sees that the spectra extend below mid-rapidity. A sum of two Gaussians \cite{maciej_cpod16} was used to fit the rapidity spectra and obtain $4\pi$ pion multiplicities which were then used to draw the {\it kink} plots (Fig.~\ref{kink_plot}).    
Figure~\ref{kink_plot} shows that the $\pi$ multiplicity, normalized to the number of wounded nucleons ($W$), increases faster with beam energy at the SPS energies in central Pb+Pb than in p+p collisions ({\it kink}). The two dependences cross at about 40$A$ GeV/c ($\sqrt{s_{NN}}=8.77$ GeV, $F=2.47$ GeV$^{1/2}$). For high SPS energies Ar+Sc follows the Pb+Pb trend and for low SPS energies Ar+Sc follows the p+p tendency. The situation is opposite for Be+Be collisions. One should note that the results suffer from model dependence of estimating $W$ (see Ref.~\cite{michal_cpod16} for details).


\begin{wrapfigure}{r}{6.cm}
\vspace{-0.6cm}
\includegraphics[scale=0.3]{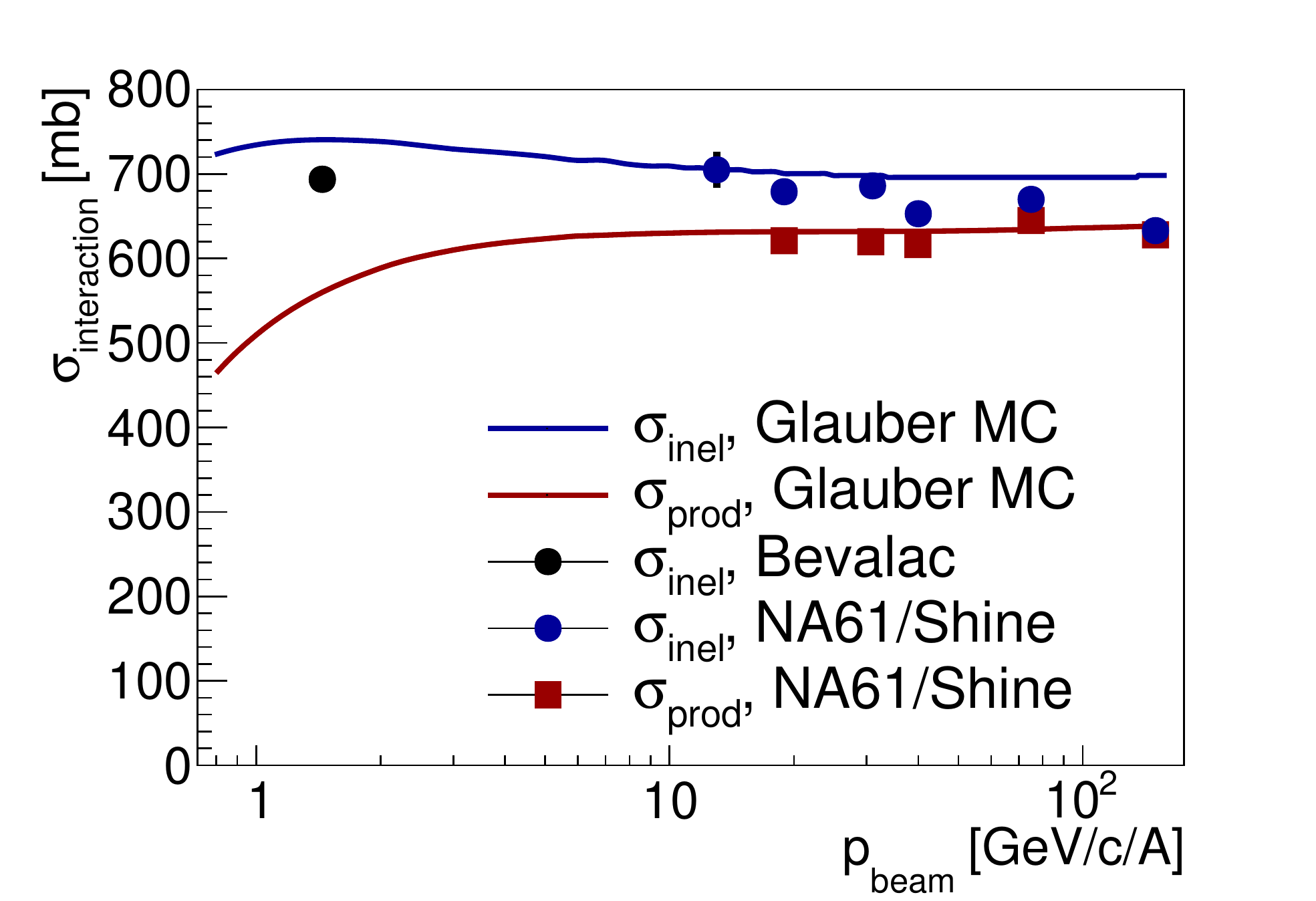}
\vspace{-0.6cm}
\caption[]{Cross-section of $^{7}$Be on $^9$Be.}
\vspace{-0.3cm}
\label{emil_cross}
\end{wrapfigure}

\section{Interaction cross-section of $^{7}$Be on $^9$Be}
Preliminary NA61 results on the interaction cross-section of $^{7}$Be on $^9$Be are presented in Fig.~\ref{emil_cross}. The values of the inelastic cross-section (initial beam particle was broken up) and the production cross section (at least one new particle in the final state) are shown as a function of beam momentum. Values measured by NA61 are in good agreement with earlier measurement at lower momentum \cite{Tanihata:1986kh}, as well as with predictions of the Glauber-based Glissando model \cite{Broniowski:2007nz}. The NA61 measurements together with $\sim$1$A$ GeV/c Bevalac data establish the energy dependence of the inelastic cross section from 1$A$ GeV to 150$A$ GeV.


\section{Search for the critical point}
The strategy of looking for the critical point (CP) of strongly interacting matter is based on a search for non-monotonic behavior of CP signatures such as fluctuations of transverse momentum, multiplicity, intermittency, etc. when the system freezes out close to the CP. NA61 uses the strongly intensive $\Delta[P_T, N]$, $\Sigma[P_T, N]$, and $\Phi_{p_T}$ measures to study transverse momentum and multiplicity fluctuations \cite{Aduszkiewicz:2015jna, Czopowicz:2015mfa}. In the Grand Canonical Ensemble they do not depend on volume and volume fluctuations. Moreover, $\Delta[P_T, N]$ and $\Sigma[P_T, N]$ have two reference values, namely they are equal to zero in case of no fluctuations and one in case of independent particle production. The recent NA61 results \cite{Czopowicz:2015mfa} show no sign of any anomaly, that can be attributed to a CP, neither in p+p nor centrality selected Be+Be collisions. On the other hand a maximum for Si+Si and C+C was observed by NA49 in transverse momentum and multiplicity fluctuations at 158A GeV/c (see Refs.~\cite{Anticic:2015fla, Czopowicz:2015mfa} and references therein). 


\subsection{Transverse momentum and multiplicity fluctuations in A+A}

\begin{figure}[ht]
\centering
\vspace{-0.39cm}
\includegraphics[width=0.24\textwidth]{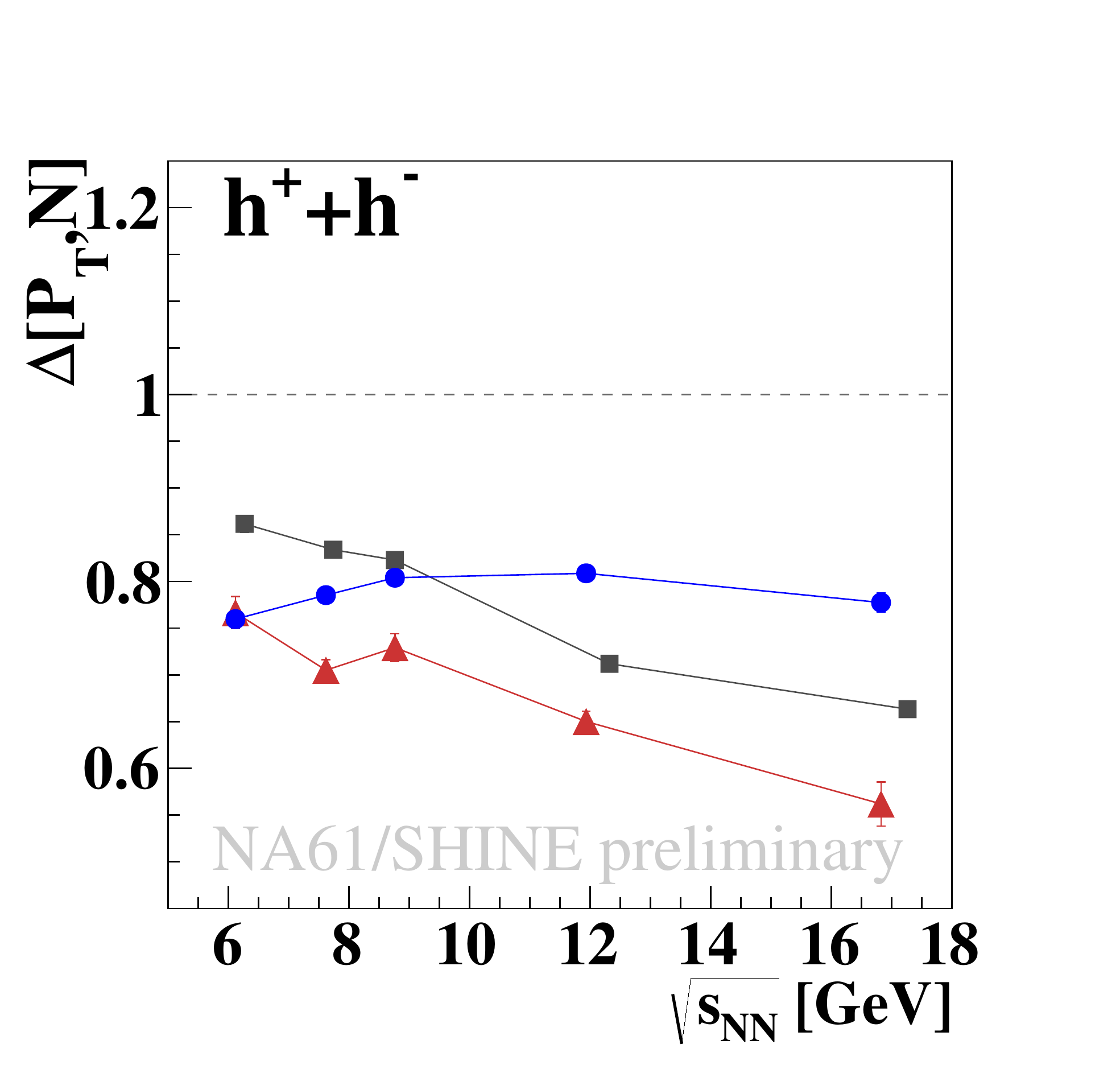}
\includegraphics[width=0.24\textwidth]{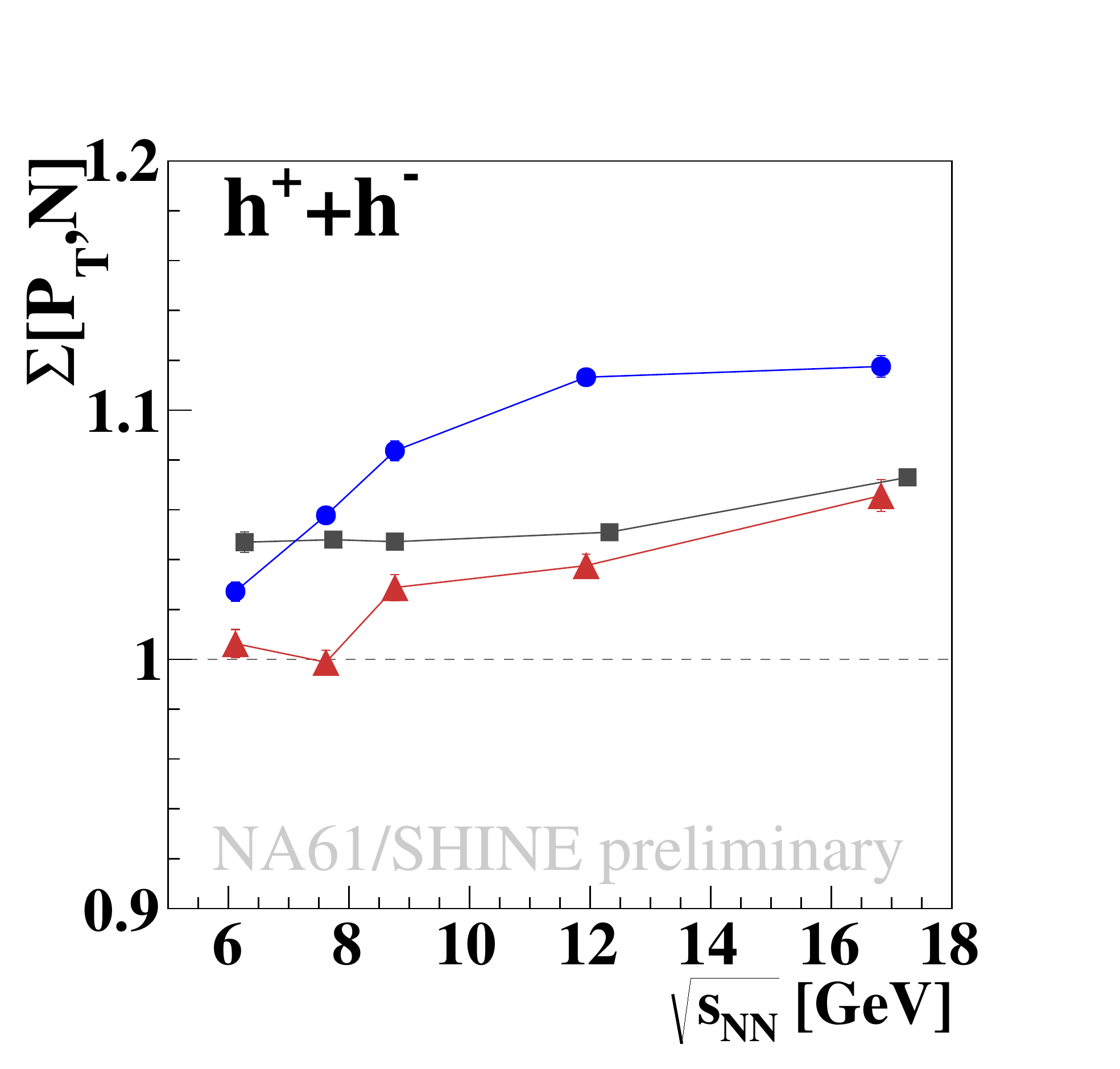}
\includegraphics[width=0.24\textwidth]{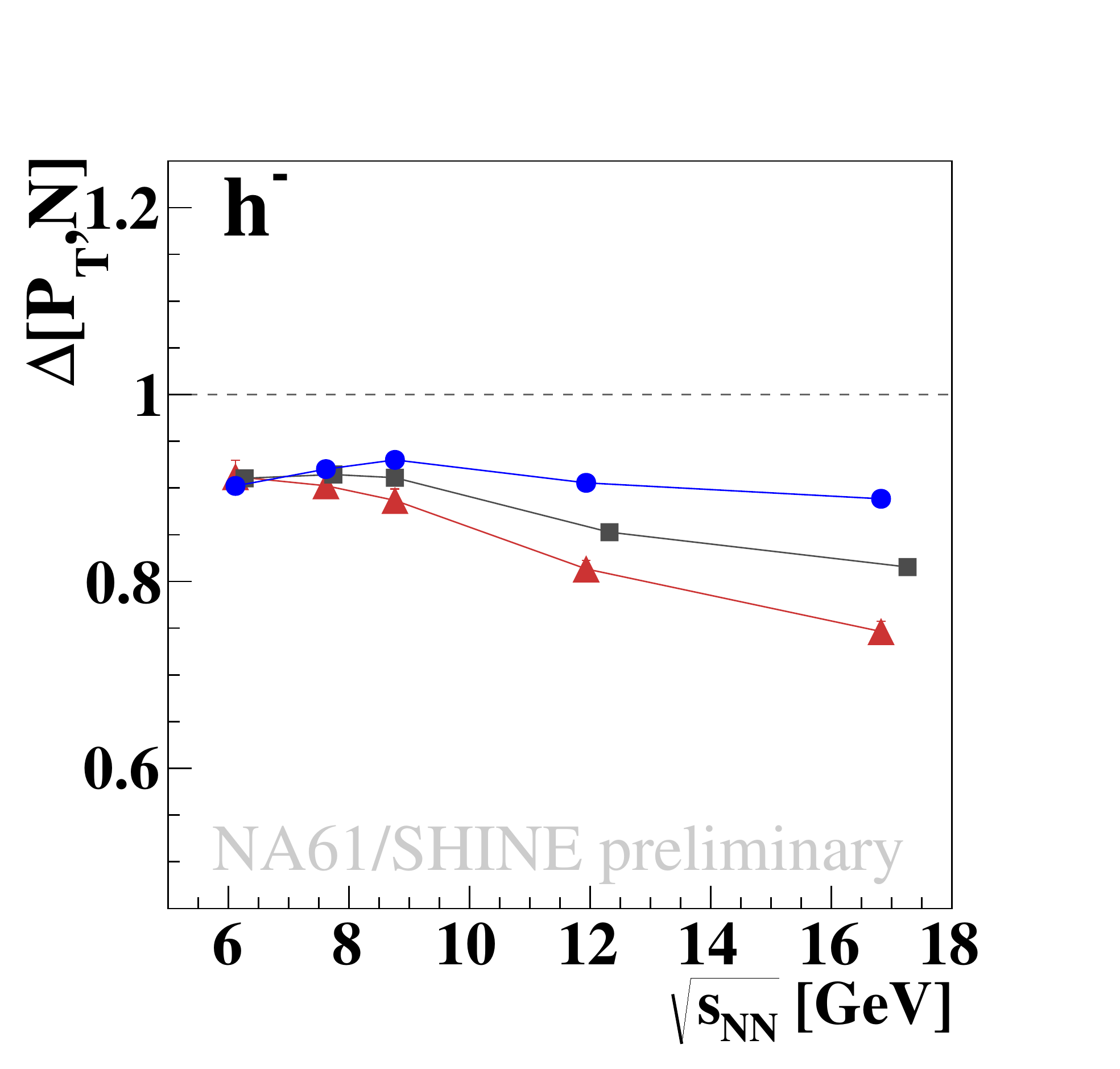}
\includegraphics[width=0.24\textwidth]{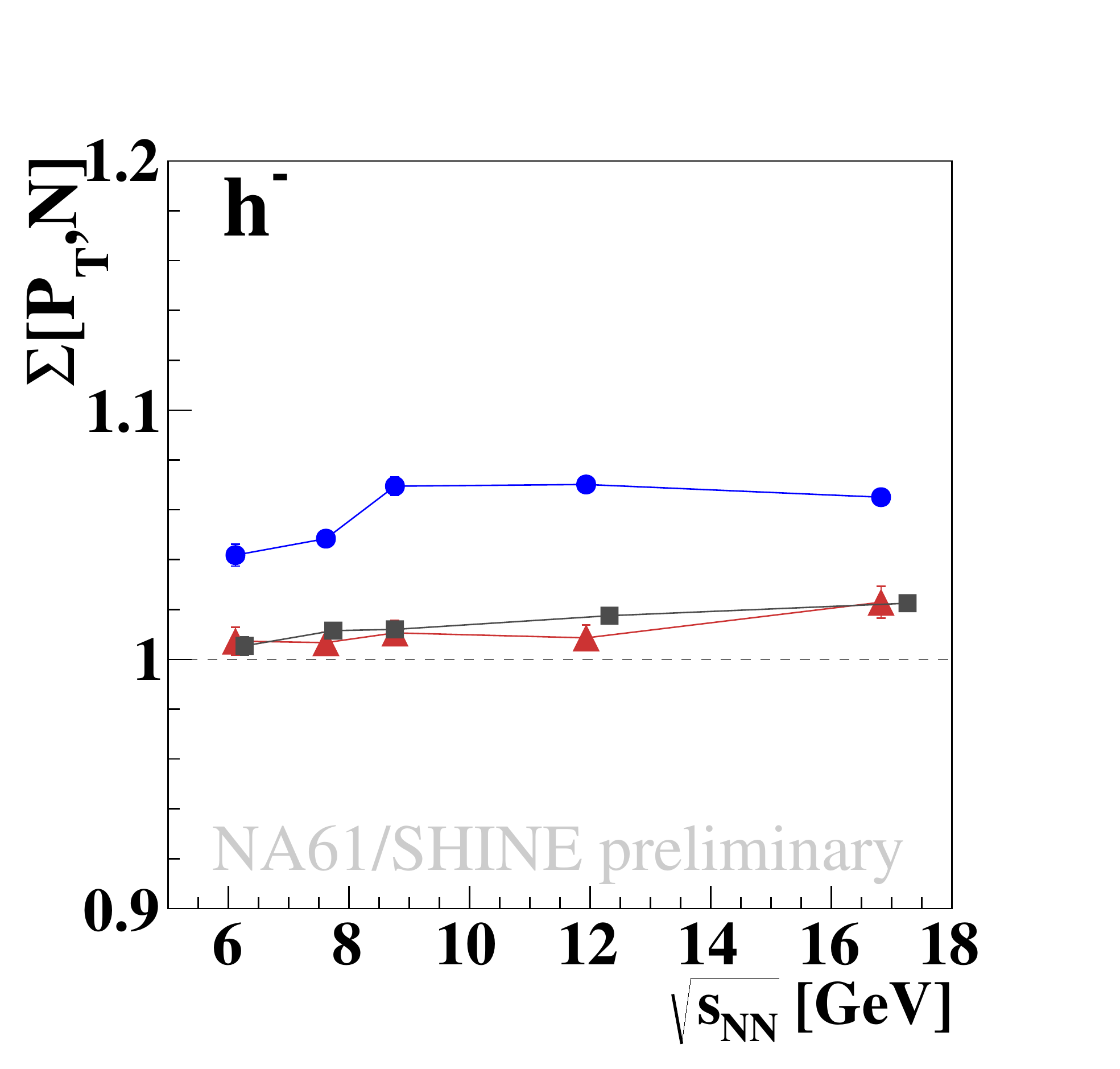}
\vspace{-0.2cm}
\caption[]{$\Delta[P_T, N]$ and $\Sigma[P_T, N]$ in inelastic p+p (grey squares), 0-5\% Be+Be (red triangles), and 0-5\% Ar+Sc (blue circles) collisions obtained by NA61 at forward-rapidity, $0< y_{\pi}< y_{beam}$, and in $p_T<1.5$ GeV/c. Results for all charged hadrons ($h^{+}+h^{-}$)
and negatively charged hadrons ($h^{-}$).
See Ref.~\cite{evgeny_cpod16} for more plots.}
\label{evgeny_pp_bebe_arsc}
\end{figure}

Figure~\ref{evgeny_pp_bebe_arsc} shows preliminary NA61 results on transverse momentum and multiplicity fluctuations in p+p, Be+Be, and Ar+Sc collisions (note that p+p and Be+Be results were already shown by NA61 \cite{Aduszkiewicz:2015jna, Czopowicz:2015mfa} but in a slightly different acceptance). These three systems show no prominent non-monotonic behavior that can be attributed to a CP. The values of $\Delta[P_T, N]$ smaller than 1 and $\Sigma[P_T, N]$ higher than 1 may be due to Bose-Einstein statistics and/or anti-correlation between event transverse momentum and multiplicity \cite{Gorenstein:2013nea}. The $\Sigma[P_T, N]$ values for $h^{-}$ at 150/158$A$ GeV/c are higher in Ar+Sc than in Be+Be and p+p, but this increase is seen also at lower energies so it is not necessarily connected with a CP.

\begin{figure}[ht]
\centering
\vspace{-0.5cm}
\includegraphics[width=0.3\textwidth]{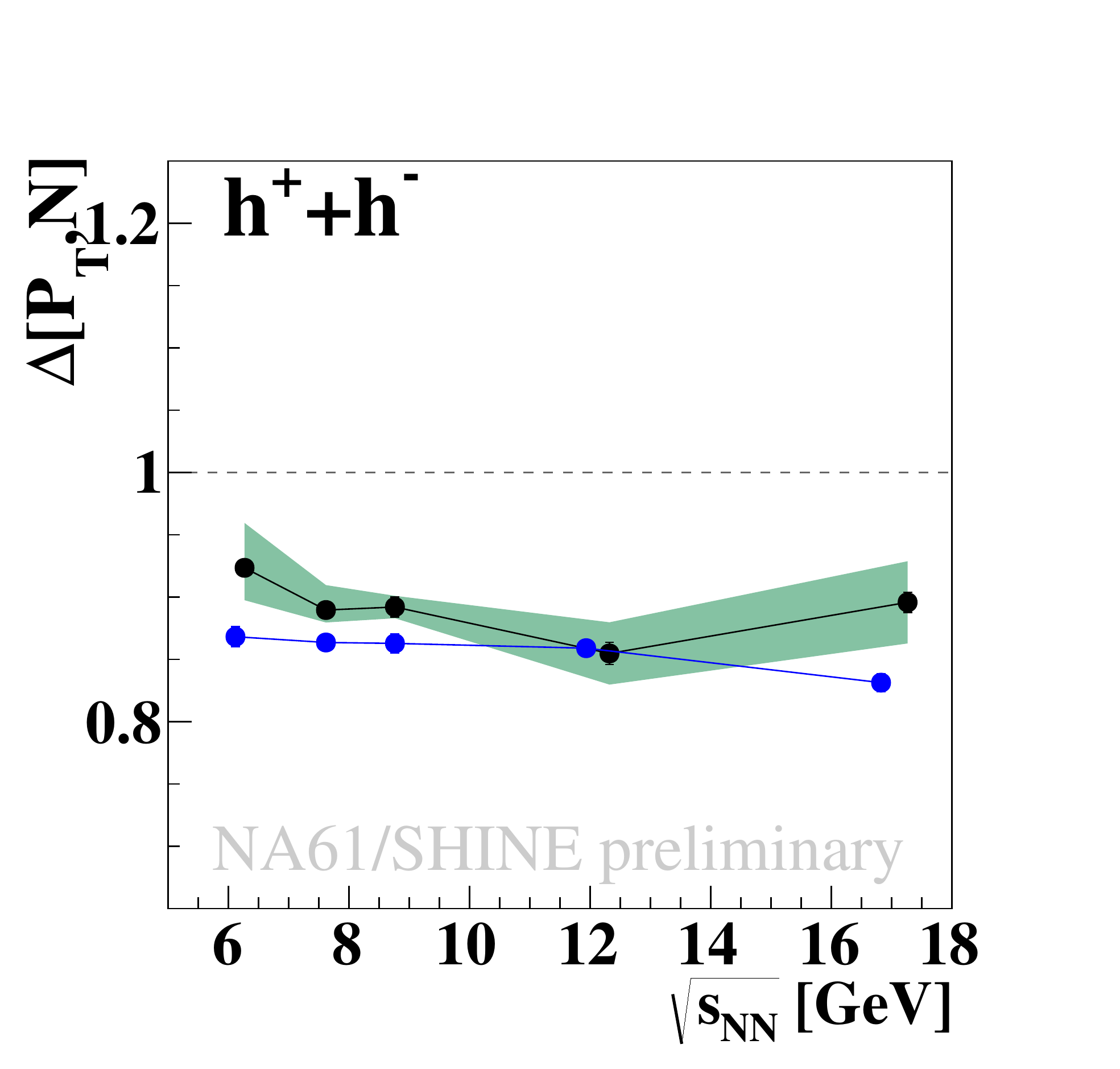}
\includegraphics[width=0.3\textwidth]{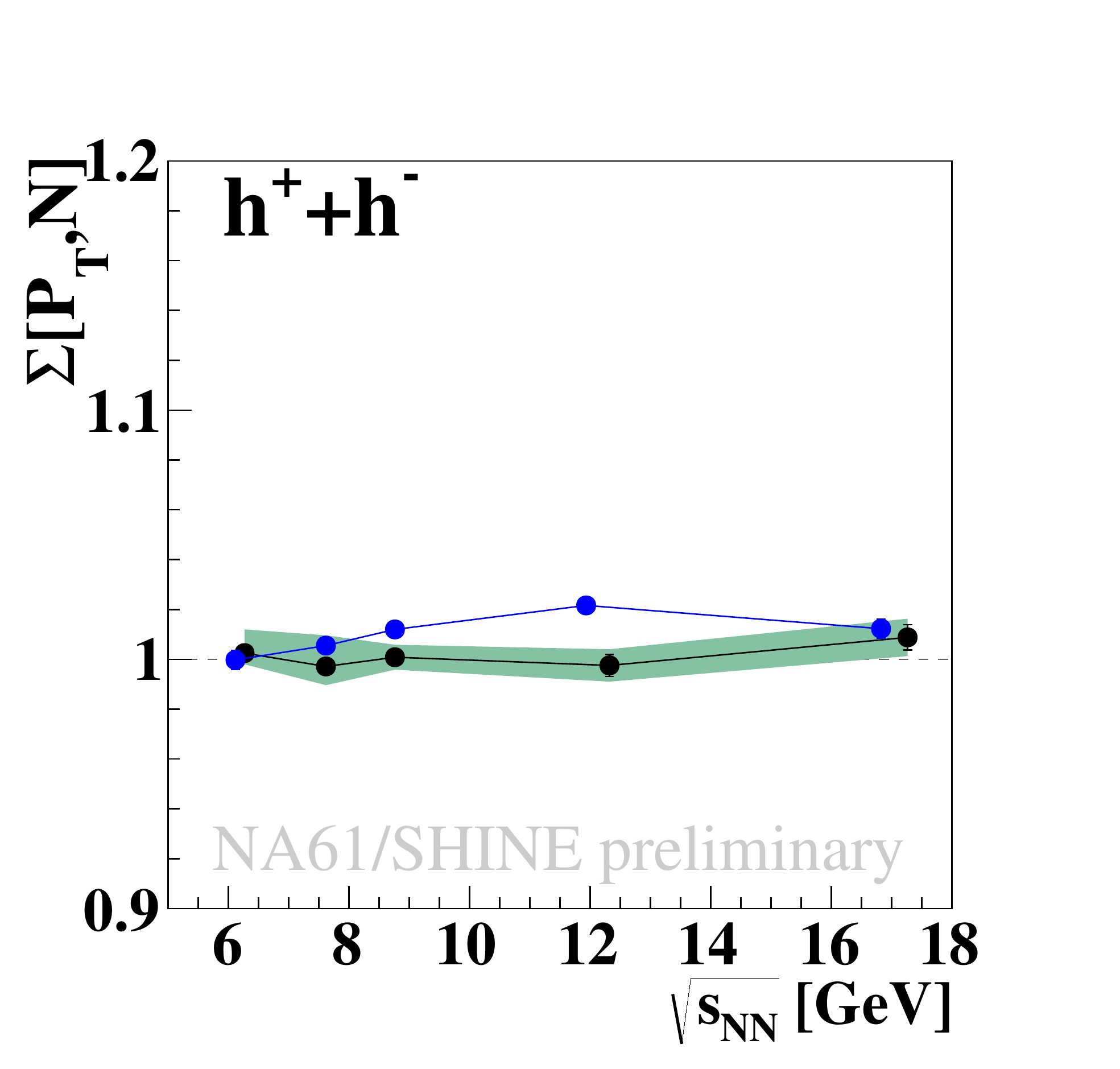}
\includegraphics[width=0.3\textwidth]{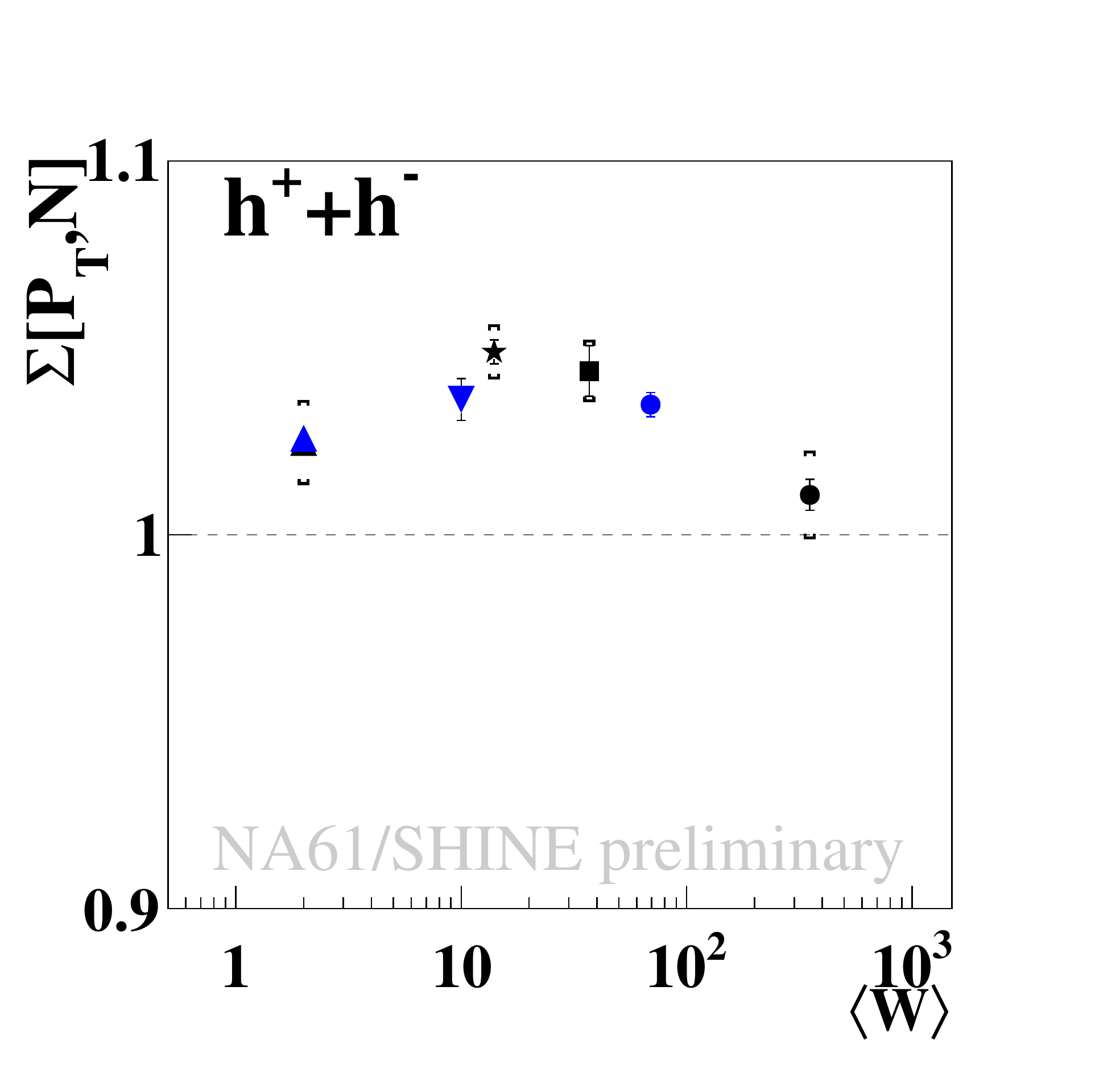}
\vspace{-0.2cm}
\caption[]{Left and middle: $\Delta[P_T, N]$ and $\Sigma[P_T, N]$ in NA61 0-5\% Ar+Sc (blue points) and NA49 0-7.2\% Pb+Pb (black points). Results in $1.1< y_{\pi}< 2.6$ and $y_{p}< y_{beam}-0.5$ with narrow azimuthal acceptance \cite{Anticic:2015fla}. Right: $\Sigma[P_T, N]$ at 150/158$A$ GeV/c. Black points are NA49 \cite{Anticic:2015fla} data (p+p, 0-15.3\% C+C, 0-12.2\% Si+Si, 0-5\% Pb+Pb), blue ones preliminary NA61 (p+p, 0-5\% Be+Be, 0-5\% Ar+Sc). Results in $1.1< y_{\pi}< 2.6$ and with almost complete azimuthal acceptance.}
\label{fluct_na49_na61}
\end{figure}

Figure~\ref{fluct_na49_na61} (left and middle) shows that NA49 Pb+Pb \cite{Anticic:2015fla} and NA61 Ar+Sc results (in NA49 narrower acceptance \cite{Anticic:2015fla}) are similar. For the system size dependence of $\Sigma[P_T, N]$ at 150/158A GeV/c (Fig.~\ref{fluct_na49_na61}, right) the NA49 \cite{Anticic:2015fla} and NA61 points show consistent trends. $\Delta[P_T, N]$ (not shown) is more centrality width sensitive \cite{Gorenstein:2013nea} and points are scattered (see Ref.~\cite{evgeny_cpod16}).

\subsection{Multiplicity fluctuations in Ar+Sc collisions}
During the conference new NA61 results on multiplicity fluctuations in Ar+Sc collisions were shown using the scaled variance of the multiplicity distribution ($\omega[N]$) and a newly defined strongly intensive $\Omega$ measure (see conference slides and Ref.~\cite{andrey_cpod16}). There is no significant non-monotonic behavior in the energy dependence of those two measures, however an interesting effect was observed. Namely, for negatively charged hadrons (the cleanest sample) at 150/158$A$ GeV/c the scaled variance of the multiplicity distribution is much below 1 for 0-0.2\% Ar+Sc and 0-1\% Pb+Pb and above 1 for p+p. This clearly shows the violation of the Wounded Nucleon Model in this fluctuation analysis. Moreover, it is also forbidden in the Ideal Boltzmann Grand Canonical Ensemble (see conference slides or Ref.~\cite{Aduszkiewicz:2015jna} for more detailed discussion). Within statistical models $\omega[N] \gg 1$, as seen in p+p, can be understood as a result of volume and/or energy fluctuations~\cite{Begun:2008fm}.


\subsection{Higher order moments of net-charge distribution in p+p collisions}
Higher order moments of multiplicity distributions (skewness $S$, kurtosis~$\kappa$) measure the non-Gaussian nature of fluctuations and are more sensitive (than the variance $\sigma^2$) to fluctuations at a CP \cite{Stephanov:2008qz, Stephanov:2011pb}. 
Moreover, they can be used to test (statistical and dynamical) models (first moments do not allow to distinguish between different types of models; already for second moments fluctuations are different in string and statistical models). 
Finally, higher moments of conserved quantum numbers ($i = B, Q, S$) allow for direct comparison to theory via susceptibilities ($S\sigma \approx \chi_i^3 / \chi_i^2$, $\kappa \sigma^2 \approx \chi_i^4 / \chi_i^2$).

\begin{figure}[ht]
\centering
\vspace{-0.4cm}
\includegraphics[width=0.32\textwidth]{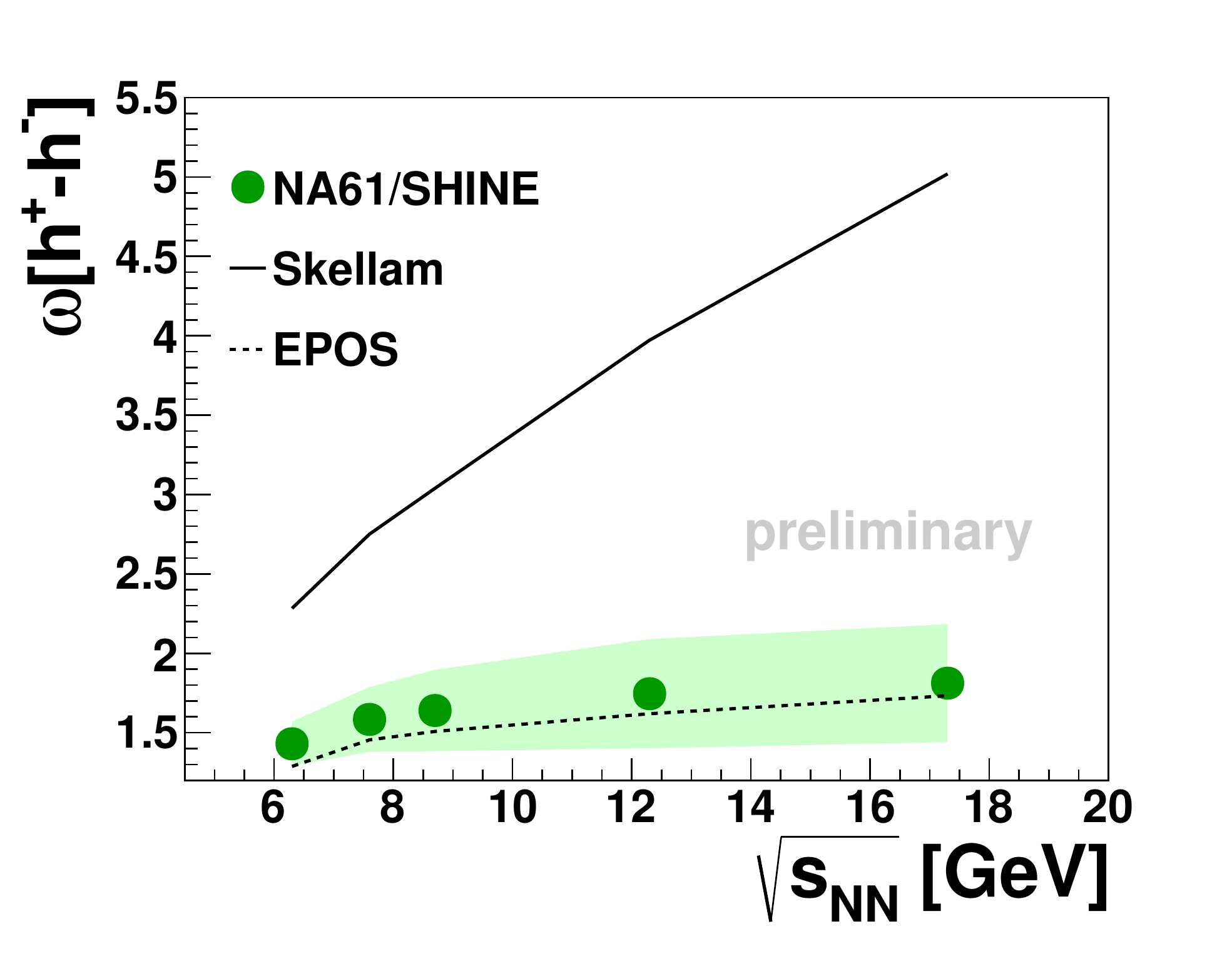}
\includegraphics[width=0.32\textwidth]{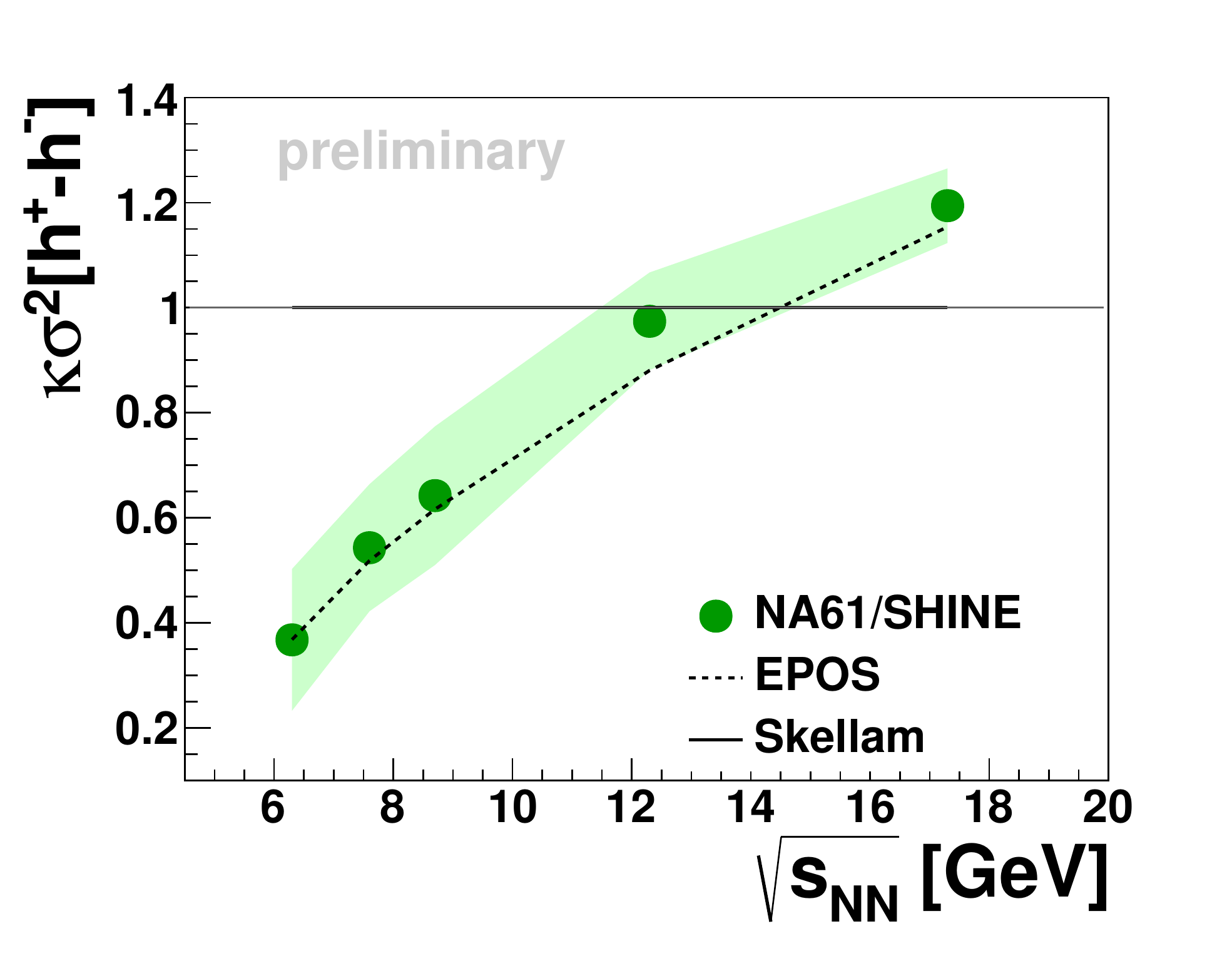}
\includegraphics[width=0.32\textwidth]{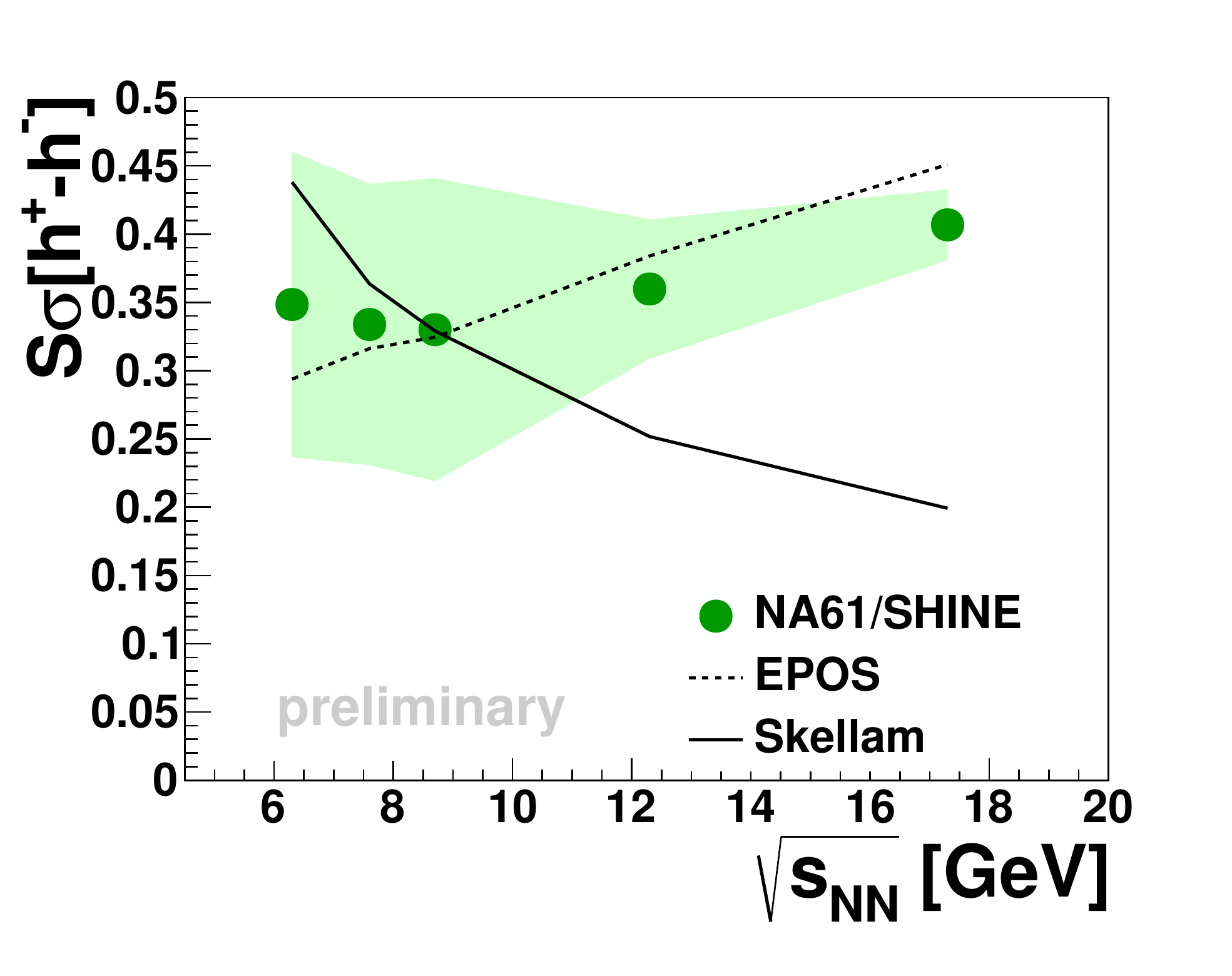}
\vspace{-0.25cm}
\caption[]{Scaled variance ($\omega$) and products of higher order moments of net-charge distributions measured in inelastic p+p interactions.}
\label{mmp_net_charge}
\end{figure}

The scaled variance ($\omega$) and products of higher order moments of net-charge distributions, measured in inelastic p+p interactions, are shown in Fig.~\ref{mmp_net_charge} (see conf. slides and Ref.~\cite{maja_cpod16} for the same quantities for $h^{-}$ only). There is no non-monotonic behavior suggesting a CP. Results do not agree with independent particle production (Skellam) but EPOS 1.99 describes the data quite well. Be+Be and Ar+Sc data will be analyzed soon.

\vspace{0.3cm}

{\footnotesize {\bf Acknowledgments:} This work was partially supported by the National Science Centre, Poland grant
2015/18/M/ST2/00125 and SPbSU research grant 11.38.242.2015.

\end{document}